\documentclass[
 reprint,
nofootinbib,
 amsmath,amssymb,
 aps,
]{revtex4-2}

\usepackage{graphicx, color}
\usepackage{dcolumn}
\usepackage{bm}
\usepackage[mathlines]{lineno}
\usepackage{amsmath}
\usepackage{amssymb}
\usepackage{bm}
\usepackage{mathrsfs}
\usepackage{float}
\usepackage{soul}
\usepackage{mathrsfs}
\usepackage{lipsum}
\usepackage{hyperref}
\usepackage[normalem]{ulem}
\hypersetup{
    colorlinks=true,       
    linkcolor=red,          
    citecolor=rp,        
    filecolor=magenta,      
    urlcolor=blue           
}
\usepackage{cleveref}
\usepackage{color}

\definecolor{rp}{cmyk}{0.2, 1, 0.6, 0}
\definecolor{rp}{cmyk}{0.2, 1, 0.6, 0}
\definecolor{green2}{cmyk}{0.27, 0, 1, 0.52}

\newcommand{\schr}{\rm Schr{\"o}dinger }

\usepackage{hyperref}
\hypersetup{%
  colorlinks = true,
  linkcolor  = rp
}

\begin{document}

\preprint{APS/123-QED}

\title{Kinetic relaxation and Bose-star formation in multicomponent dark matter- I}

\author{Mudit Jain}
\email{mudit.jain@rice.edu}

\author{Mustafa A. Amin}
\email{mustafa.a.amin@rice.edu}
\author{Jonathan Thomas}
\email{jt57@rice.edu}
\author{Wisha Wanichwecharungruang}
\email{wisha@rice.edu}

\affiliation{Department of Physics and Astronomy, Rice University, Houston, Texas 77005, U.S.A.}

\date{\today}

\begin{abstract}
Using wave kinetics, we estimate the emergence time-scale of gravitating Bose-Einstein condensates/Bose stars in the kinetic regime for a general multicomponent Schr\"{o}dinger-Poisson (SP) system. We identify some effects of the diffusion and friction pieces in the wave-kinetic Boltzmann equation (at leading order in perturbation theory) and  provide  estimates for the kinetic nucleation rate of condensates. We test our analysis using full $3+1$ dimensional simulations of multicomponent SP system.
With an eye towards applications to  multicomponent dark matter, we investigate two general cases in detail. First is a massive spin-$s$ field with $N=2s+1$ components (scalar $s=0$, vector $s=1$ and tensor $s=2$). We find that for a democratic population of different components, the condensation time-scale is $\tau_{(s)}\approx \tau_0\times N$, where $\tau_0$ is the condensation time scale for the scalar case. Second is the case of two scalars with different boson masses. In this case, we map-out how the condensation time depends on the ratios of their average mass densities and boson masses, revealing competition and assistance between components, and a guide towards which component condenses first. For instance, with $m_1 < m_2$ and not too disparate mass densities, we verify that the time scale of condensation of the first species quickly becomes independent of $m_2/m_1$, whereas for equal average number densities, the emergence time scale decreases with increasing $m_2/m_1$.
\end{abstract}

\maketitle


\section{Introduction}\label{sec:intro}

Sufficiently light bosonic dark matter leads to a plethora of wave phenomenon (see~\cite{Ferreira:2020fam,Hui:2021tkt} for a recent reviews), including the condensation of Bose stars  in the kinetic regime via gravitational interactions. In an elegant paper~\cite{Levkov:2018kau}, Levkov, Panin and Tkachev provide numerical simulations and an analytic estimate of the condensation time-scale in the case of a single scalar field. Also see~\cite{Eggemeier:2019jsu,Chen:2020cef,Hertzberg:2020hsz,Chan:2022bkz} for related recent analyses and some applications to astrophysical settings. These analyses were carried out using a single component non-relativistic Schr\"{o}dinger-Poisson system.

In this paper, we investigate kinetic condensation in a multicomponent Schr\"{o}dinger-Poisson system,  where each component can have equal or different boson mass and mass density, and explore the nature of nucleated Boson stars. Such multicomponent SP systems naturally describe $2s+1$ component spin-$s$ bosonic dark matter ($s=1$ for vector and $s=2$ for tensor dark matter), or when dark matter consists of a collection of scalar fields.

Non-relativistic Bose stars/solitons in spin-$s$ fields, where $s>0$, have been recently studied in the literature \cite{Aoki:2017ixz,Adshead:2021kvl,Jain:2021pnk}. Such solitons can carry macroscopic intrinsic spin angular momentum~\cite{Jain:2021pnk} (unlike ``hedgehog"-like Proca stars~\cite{Brito:2015pxa}), which can in turn lead to novel observational effects~\cite{March-Russell:2022zll, Amin:2023imi}. The $s=0$ case has of course been explored for several decades~\cite{Ruffini:1969qy} (see~\cite{Chavanis:2022fvh} for a review). For $s=1$, the solitons have been seen to form due to gravitational interactions from cosmological initial conditions~\cite{Gorghetto:2022sue}, and also from mergers of halos/solitons~\cite{Amin:2022pzv}. However, their emergence via condensation in the kinetic regime has not been explored before. Similarly, solitons in dark matter made up of multiple scalar fields (with different, but comparable, boson masses), have been investigated in the literature, especially in the context of core profiles~\cite{Guo:2020tla,Huang:2022ffc,Street:2022nib,Glennon:2023jsp}. However, their formation via kinetic relaxation has not been investigated. We hope that our work sheds light on this subject, and will be useful for exploring their observational implications. 

Starting with the multicomponent SP system, we derive the wave-kinetic / Boltzmann equation valid in the kinetic regime. Under an eikonal approximation (small scattering angle limit), the system simplifies considerably which upon re-writing in the Fokker-Planck form, reveals the diffusion and friction terms. For the purposes of condensate nucleation, we focus on the behavior of the distribution function at vanishing momenta. We provide a set of coupled ordinary differential equations for their evolution, and also estimate an initial condition based condensation rate. 

As a consequence, we find that for a spin-$s$ system with $N=2s+1$ components (necessarily with equal boson masses for each component), and with statistically equivalent initial conditions, the time scale of condensation scales with the number of components. On the other hand for a two component system, with potentially different mass densities and boson masses, we map-out the landscape  of condensation times, revealing for example, the regime when condensation time becomes independent of the ratio of boson masses and when condensation times are determined by the heavier or lighter component.

We carry out a suite ($\sim 100$) of 3-dimensional numerical simulations of the multicomponent SP system to explore the domain of validity of our estimates. We find that the results are in general agreement with the analytic estimates.

The rest of the paper is organized as follows. In Sec.~\ref{sec:model} we describe the general model of multicomponent dark matter with only gravitational self-interactions. Leaving details of the derivation of multicomponent wave kinetic equation for appendix~\ref{app:wave_kinetic_derivation}, and its subsequent reduction in the eikonal approximation for appendix~\ref{sec:app_eikonal_approx}, in Sec.~\ref{sec:kin_relax} we discuss the general structure of the Boltzmann / Fokker-Planck equation for our multicomponent SP system. We provide estimates of the rate of change of distribution functions at vanishing momenta, which are relevant for the nucleation time scales of gravitating condensates. In subsequent subsections~\ref{subsection:spin-s} and~\ref{subsection:unequal-masses}, we specialize to the two cases of interest mentioned above, discuss the simulation results, and provide comparisons with analytical estimates. Finally, in \ref{sec:summary}, we summarize our work. Details of numerical simulations are provided in yet another appendix~\ref{app:numerical_details}.

\noindent{{\bf Conventions}}: Unless stated otherwise, we will work in the units where $\hbar = c = 1$.
\section{Model}\label{sec:model}
We are interested in sufficiently subhorizon dynamics, and hence ignore Hubble expansion. In this case, the dynamics of the multicomponent dark matter field is described by the following non-relativistic Schr\"{o}dinger-Poisson (SP) system of equations:
\begin{align}\label{eq:SP_general}
    i\frac{\partial}{\partial t}\psi_a &= -\frac{1}{2m_a}\nabla^2\psi_a + m_a\Phi\,\psi_a\nonumber\\
    {\rm where} \qquad \nabla^2\Phi &= 4\pi G\sum_{b}m_b\,\psi^{\ast}_b \psi_b.
\end{align}
If $m_a=m$ for all $``a"$, then $\psi_a$ can be thought of as components of a spin-$s$ field. Here, $``a"$ ranges from $1$ to $N=2s+1$. In this case, the above system has a $U(2s+1)$ symmetry, leading to conservation of extra charges (apart from mass conservation within each component) such as iso-spin and/or spin~\cite{Jain:2021pnk}.

More generally, each component $\psi_a$ can have a different mass, in which case each component represents a collection of scalar particles (distinct from other components). Correspondingly, owing to a separate $U(1)$ symmetry in each scalar sector, the total number of particles within each sector is conserved. 

We are interested in kinetic relaxation/condensation. In the kinetic regime, the time-scales of interactions are much longer than the oscillation time of the free waves. In addition, the wavelengths are  much smaller than the size of the system under consideration. Physically, this translates to having the dark matter halo size much larger than the de-Broglie scale for the dark matter field.

\section{Kinetic relaxation}\label{sec:kin_relax}

A formal estimate for the time-scale of Bose-Einstein condensation in the kinetic regime may be obtained by means of the wave kinetic equation. While we derive a general multicomponent wave kinetic equation (with arbitrary $2$ body interaction) using a random phase approximation in appendix~\ref{app:wave_kinetic_derivation}, for our purposes in the present paper we are only interested in gravitational interactions. In this case, the wave kinetic equation for the occupation number function $f^a_{\bm k/m_a} = |\Psi^a_{\bm k/m_a}|^2$ for species $``a"$, takes the following form
\begin{widetext}
    \begin{align}\label{eq:Boltzmann_gravity}
    &\frac{\partial f^a_{\bm k/m_a}}{\partial t} = \sum_{b}\int\frac{\mathrm{d}{\bm p}}{(2\pi)^3}\,\mathrm{d}\sigma_{{\bm k}_a + {\bm p}_b \rightarrow {\bm q}_b + {\bm \ell}_a}\,|{\bm v}_a - \tilde{\bm v}_b|\,\Biggl[(f^a_{\bm k/m_a} + f^b_{\bm p/m_b})f^a_{\bm \ell/m_a}f^b_{\bm q/m_b} - (f^a_{\bm \ell/m_a} + f^b_{\bm q/m_b})f^a_{\bm k/m_a}f^b_{\bm p/m_b}\Biggr]\,,
    \nonumber\\
    &{\rm where} \quad \mathrm{d}\sigma_{{\bm k}_a + {\bm p}_b \rightarrow {\bm q}_b + {\bm \ell}_a} = \frac{\mathrm{d}{\bm q}}{(2\pi)^3}\frac{\mathrm{d}{\bm \ell}}{(2\pi)^3}\frac{1}{|{\bm v}_a - \tilde{\bm v}_b|}\frac{(4\pi G m_a m_b)^2}{|{\bm k}-{\bm \ell}|^2}\Biggl(\frac{1}{|{\bm k}-{\bm \ell}|^2} + \frac{\delta_{ab}}{|{\bm k}-{\bm q}|^2}\Biggr)\,\times\nonumber\\
    &\qquad\qquad\qquad\qquad\qquad\qquad (2\pi)^4\,\delta^{(3)}(\bm k + \bm p - \bm q - \bm \ell)\,\delta(E^a_{\bm k} + E^b_{\bm p} - E^b_{\bm q} - E^a_{\bm \ell})\,.
    \end{align}
\end{widetext}
Here ${\bm v}_a$ and $\tilde{\bm v}_b$ are incoming``velocities" for the species $``a"$ and $``b"$ carrying momentum ${\bm k} = m_a{\bm v}_a$ and ${\bm p} = m_b\tilde{\bm v}_b$ respectively, and $\bar{\rho}_c = m_c(2\pi)^{-3}\int\mathrm{d}{\bm k}\,f^c_{\bm k}$ is the average mass density for any $c^{\rm th}$ species. Also, $E^{k}_{a} = k^2/2m_a$ is the free wave dispersion relation, and the quantity $\mathrm{d}\sigma_{{\bm k}_a + {\bm p}_b \rightarrow {\bm q}_b + {\bm \ell}_a}$ is the differential cross section for the process ${\bm k}_a + {\bm p}_b \rightarrow {\bm q}_b + {\bm \ell}_a$. The summation over $``b"$ simply reflects the fact that any species $``a"$ gravitationally interacts with all the other species (including species $``a"$ itself), and can be readily contrasted with a single species/scalar case. Also note the term $\propto \delta_{ab}$ in the differential cross-section, which can be readily interpreted as an interference between the $u$ and $t$ interaction channels.\footnote{This interference term gives negligible contribution in the eikonal/small-angle approximation (relevant for long range interactions), but could become important for other (e.g. short range) interactions. See appendix~\ref{sec:app_eikonal_approx} for details.} Furthermore, the above wave-kinetic equation can be contrasted with its ``non-wavelike" counterpart (i.e. the usual kinetic equation for point like particles): The bracket terms carrying the sum of occupation number functions are simply unity in the latter case.\\

In general, on account of interactions, waves exchange energy and the occupation number function evolves with the characteristic time of this evolution being $\sim (\partial\log f/\partial t)^{-1}$ (for every species). As a result, an important phenomenon of `condensation' can occur. As we shall see explicitly for the case of gravity, the occupation number function for the condensing species develops an increasing support over smaller ${\bm k}$ values. Once enough support is developed, the gravitational potential energy of such waves becomes capable of balancing their own gradient pressure within a region, hence the emergence/nucleation of a soliton like object.\footnote{Note that in general, the existence of a \textit{spatially localized} condensate relies on there being an attractive interaction that can counterbalance the gradient pressure (and/or repulsive self interaction). See for example~\cite{Guth:2014hsa} for an analysis.} In order to make analytical progress for the estimation of this condensation rate, we work with an eikonal approximation where the change in relative velocities of the outgoing waves in assumed to be small (as compared to the relative velocities of the incoming waves). Leaving a detailed calculation for appendix~\ref{sec:app_eikonal_approx}, the wave-kinetic Boltzmann equation reduces to the following Fokker-Planck form at leading order perturbation theory:
\begin{widetext}
\begin{align}\label{eq:Boltzmann_gravity_eikonal}
    \frac{\partial f^a_{{\bm v}_a}}{\partial t}
    &= \sum_{b}m_b^3\frac{\Lambda}{4\pi}\frac{(4\pi m_am_bG)^2}{m_a}\nabla_{{v}^i_a}\Biggl[\frac{\mathcal{D}^{ab}_{ij}}{2m_a}\nabla_{v^j_a}f^a_{{\bm v}_a} + \frac{\mathcal{F}^{ab}_i}{m_b}f^a_{{\bm v}_a}\Biggr]\nonumber\\
    {\rm where} &\qquad \mathcal{D}^{ab}_{ij} = \int\frac{\mathrm{d}\tilde{\bm v}_b}{(2\pi)^3}\,f^b_{\tilde{\bm v}_b}\,\frac{\delta_{ij} - \hat{u}_i\hat{u}_j }{u}\,f^b_{\tilde{\bm v}_b}\quad{\rm and}\quad \mathcal{F}^{ab}_i = f^a_{{\bm v}_a}\int\frac{\mathrm{d}\tilde{\bm v}_b}{(2\pi)^3}\,\frac{\hat{u}_i}{u^2}\,f^b_{\tilde{\bm v}_b}\,,\quad{\rm with}\quad {\bm u} = {\bm v}_a - \tilde{\bm v}_b
\end{align}
\end{widetext}
Here, we have relabelled the occupation number functions using ``velocity" vectors, with ${\bm v}_a = {\bm k}_a/m_a$ being the incoming velocity vector for the $a$ species, and $\tilde{{\bm v}}_b$ being the velocity vector for the incoming $b$ species, giving ${\bm u} = {\bm v}_a - \tilde{\bm v}_b$ as the relative velocity between the two. Also, $\Lambda$ is the Coulomb logarithm (see~\ref{sec:app_eikonal_approx} for details). Equation~\eqref{eq:Boltzmann_gravity_eikonal} is our master Boltzmann equation (under the small angle approximation) which dictates the evolution of the occupation number functions.\footnote{The wave-kinetic equation differs from the usual (non-wavelike/particle) counterpart: the extra factors of $f^b_{\tilde{\bm v}_{b}}$ and $f^a_{{\bm v}_a}$ in the diffusion and friction coefficients are absent in the latter.}\\

The two terms on the right hand side of the Fokker-Planck equation~\eqref{eq:Boltzmann_gravity_eikonal} are conveniently understood by means of the  (velocity dependent) diffusion and friction coefficients $\mathcal{D}^{ab}_{ij}$ and $\mathcal{F}^{ab}_i$ respectively. 
For an interaction of wave type $``a"$ with wave type $``b"$, a physical effect of the diffusion term is to decrease the occupation number function $f^a_{{\bm v}_a}$ at places where it is convex, while increasing it at places where it is concave (in the plane perpendicular to ${\bm u}$, with `sheer stress' of the form $\sim 1/u$). On the other hand, an effect of the friction term is to enhance $f^a_{{\bm v}_a}$ due to the `friction force' $\sim 1/u^2$ being directed towards ${\bm v}_a$. Specifically, $\nabla_{v^i_a}\mathcal{F}_i^{ab}$ includes $4\pi f^{a}_{{\bm v}_a}f^b_{{\bm v}_a}/(2\pi)^3$, which together with the factor of $f^a_{{\bm v}_a}/m_b$ may be regarded as a positive definite source term for the evolution of $f^{a}_{{\bm v}_a}$. This heuristic understanding is similar to the non-wavelike/particle like case, albeit with the crucial difference of there being extra factors of $f^b_{\tilde{\bm v}_{b}}$ and $f^a_{{\bm v}_a}$ in the diffusion and friction terms respectively due to wave dynamics. These extra terms, sometimes referred to as Bose enhancement factors, have an important role to play in nucleation of condensates.\\ 

We note that the above understanding of these effects of the diffusion and  friction terms, and a subsequent nucleation of a condensate is reflected in a preliminary calculation of moments of the distribution function $f^a_{{\bm v}_a}$. For instance even for a single species case, assuming a Gaussian initial {\it ansatz} for the distribution function (c.f. Eq.~\eqref{eq:Gaussian_initialansatz} ahead), we calculate the rate of change of different moments at the initial instant. We find that while $d\langle v_a\rangle/dt|_{t=0} < 0$, $d\langle v_a^n\rangle/dt|_{t=0} > 0$ for $n \geq 3$, with $d\langle v_a^2\rangle/dt|_{t=0} = 0$ being the boundary case. This indicates that the evolution of $f_{v}$ is such that it tries to break into a condensate part where the friction dominates over diffusion (developing increasing support towards smaller velocities), and a remaining part where this may not be true.\\

For the purposes of condensate/soliton nucleation within any species $``a"$, we may therefore focus on the behavior of its occupation number function at small velocities, i.e. the quantity $\lim_{{\bm v}_a \rightarrow 0}\frac{\partial f^a_{{\bm v}_a}}{\partial t}$, due to all the other species (including itself) in the bath. (We of course do not make the same assumption about the species being integrated over.) We assume homogeneity and isotropy (until the nucleation of the condensate) along with an assumption of quadratic functional dependence of occupation number functions at small velocities. Under these assumptions, the diffusion piece $\nabla_{v^i_a}\nabla_{v^j_a} f^{a}_{{\bm v}_a}|_{v_a \rightarrow 0} \rightarrow -\tilde{\beta}_a\delta_{ij}\,f^a_{0}/\sigma_a^2$, giving the subsequent velocity integral to be $\mathcal{D}^{ab}_{ij}\delta_{ij} \rightarrow 2\times2\pi\sigma_b^2(\bar{\rho}_b^2/m_b^8\sigma_b^6)\beta'_b$. Here, $\sigma_b$ characterizes the initial Gaussian width of the distributions, and $\bar{\rho}_b$ is the spatially averaged mass density of species $b$. Also, $\tilde{\beta}_b$ parameterizes deviations from gaussianity of the ratio of the curvature of $f^b_{0}$ versus $f^b_{0}$ (measured in units of $\sigma_a$),
while $\beta'_b$ characterizes deviations from gaussianity of the full integral in $\mathcal{D}^{ab}_{ij}$.\footnote{While in general time dependent, we expect the time variation of both $\beta'_b$ and $\tilde{\beta}_b$ to not be too significant throughout most of the evolution of the occupation number functions before the nucleation of condensates.} 
For the relevant piece in the friction term, we simply have $(\nabla_{v^i_a}\mathcal{F}^{ab}_i)f^a_{{\bm v}_a} \rightarrow 4\pi (2\pi)^{-3}f^b_{0}(f^a_0)^2$. Furthermore, to extract overall scalings of the distribution function $f^b_0$, we define a function $g_b(t)$ such that
\begin{equation}
f^b_0(t)\equiv (2\pi)^{3/2}\frac{\bar{\rho}_b}{m_b^4\sigma_b^3}\times g_b(t)
\end{equation}
where $g_b(t)$ carries all the time-dependence of the distribution function near small velocities, with $g_b(t=0)=1$. With these replacements, we finally arrive at the following
\begin{align}\label{eq:f0_evolveDE}
    \dot{g}_a &= \sum_{b}\frac{\Lambda_b\,(4\pi G)^2\bar{\rho}}{2\sigma_a^3\sigma_b^3}\Biggl[2\frac{\bar{\rho}_a}{m_a^3}g_ag_b - \beta_{ab}\frac{\bar{\rho}_b\sigma_a}{m_b^3\sigma_b}\Biggr]g_a\,,
\end{align}
where we have combined $\tilde{\beta}_a$ and $\beta'_b$ into a single $\beta_{ab}$.
As a quick exercise for a single species, we can solve this differential equation and take the time when $g$ changes significantly, as an estimate for the nucleation time of the condensate. Denoting $\tau_{\rm gr} \equiv 2m^3\sigma^6/(\Lambda(4\pi G)^2\bar{\rho}^2)$, we get $\dot{g} = \tau^{-1}_{\rm gr}(2g^3-\beta g)$, which gives $\tau_0 \sim \tau_{\rm gr}\,\log(2/(2-\beta))/(2\beta)$ under the assumption of $\beta = \rm{const.}$ (and where $\tau_0$ is the time when $g \rightarrow \infty$).

For concreteness, we also evaluate the above rate of change at the initial instant $\Gamma_a\equiv d\log g_a/dt|_{t=0}$:
\begin{align}\label{eq:Rate_initial}
    \Gamma_a = \sum_{b}\frac{\Lambda\,(4\pi G)^2\bar{\rho}_b}{2\sigma_a^3\sigma_b^3}\Biggl[2\frac{\bar{\rho}_a}{m_a^3} - \beta_{ab}\frac{\bar{\rho}_b\sigma_a}{m_b^3\sigma_b}\Biggr]\,,
\end{align}
where the $\beta$ parameters are simply informed by the initial condition, and take this as an estimate for the rate of condensate nucleation. The corresponding time of course being $\tau_a \sim \Gamma_a^{-1}$.\footnote{Note that for a Gaussian initial ansatz~\eqref{eq:Gaussian_initialansatz}, $\beta_{ab} = 1$ at the initial instant.}
Once again, for a single component case with initial condition~\eqref{eq:Gaussian_initialansatz}, we get $\tau_0 \sim \tau_{\rm gr}$.

From the above estimates for the single component case, the condensation time {\it scales} with relevant parameters similar to~\cite{Levkov:2018kau}, but the numerical factors are not identical. Our estimate is based on using Gaussian initial conditions to calculate the right-hand side of~\eqref{eq:Boltzmann_gravity_eikonal} explicitly, near vanishing momenta. To the best of our understanding, authors in~\cite{Levkov:2018kau} replace derivatives, integration measures, relative velocities and occupation number functions with respective scalings in Eq.~\eqref{eq:Boltzmann_gravity_eikonal} (specialized to a single component). They then fit an order unity co-efficient which depends on initial conditions from simulations. We thus expect the scalings to match, but not the explicit numerical factors. With multiple species, however, the scaling with densities, boson masses and initial velocity dispersions becomes non-trivial and one needs to keep track of differences arising from the friction and diffusion terms.\\

Before moving on, we would like to caution the reader that Eqs.~\eqref{eq:f0_evolveDE} and~\eqref{eq:Rate_initial} are not the most general equations that capture behavior of \textit{any} distribution function $f$ at vanishing momenta, at all times and at the initial instant respectively. They only apply in so far as the leading dependence of $f$ on momenta is quadratic (at small momenta). On the contrary, the Boltzmann equation~\eqref{eq:Boltzmann_gravity_eikonal} of course contains all the necessary details (in the leading order perturbation theory).\\ 

For simulations, in this paper we shall focus on two different scenarios. First, we will consider a spin-$s$ field with $N = 2s+1$ components, with the boson mass for each component being equal. The other case would be the opposite scenario where the different components are simply scalar fields and therefore have naturally different masses. For example this could be the case of dark matter comprising of Axiverse axions~\cite{Arvanitaki:2009fg}. For this multi-scalar case, we shall only consider the two-component case in detail. Next, owing to violent relaxation in the physical case of dark matter physics, we shall assume that all the components have the same characteristic velocity. For simulation purposes, we numerically evolve the SP system~\eqref{eq:SP_general}, with 
the following initial distribution/occupation number function for every $a^{\rm th}$ species\footnote{Note that the initial conditions used by~\cite{Levkov:2018kau}, for the scalar $s=0$ case, differs by $\sigma\rightarrow \sigma/\sqrt{2}$. Also note that while we do not discuss  initial conditions that are Dirac-Delta functions in velocity space at finite $\sigma$ (as investigated by~\cite{Levkov:2018kau}), we briefly mention what we see in some sample simulations in appendix~\ref{app:numerical_details}, and how it relates to the discussion in this section.}
\begin{align}\label{eq:Gaussian_initialansatz}
    f^a_{\bm v_a}\Bigl|_{t = 0} = |\Psi^a_{{\bm k}/m_a}|^2\Bigl|_{t=0} = \frac{(2\pi)^{3/2}\bar{\rho}_a}{m_a(m_a \sigma_a)^3}\,e^{-\frac{v_a^2}{2\sigma_a^2}}\,,
\end{align}
with $\sigma_a = \sigma$ for every species, and random phases for every wavenumber (for each species). The details of the initial conditions are provided in appendix~\ref{app:numerical_details}.

\begin{figure}
    \centering
    \includegraphics[width=0.5\textwidth]{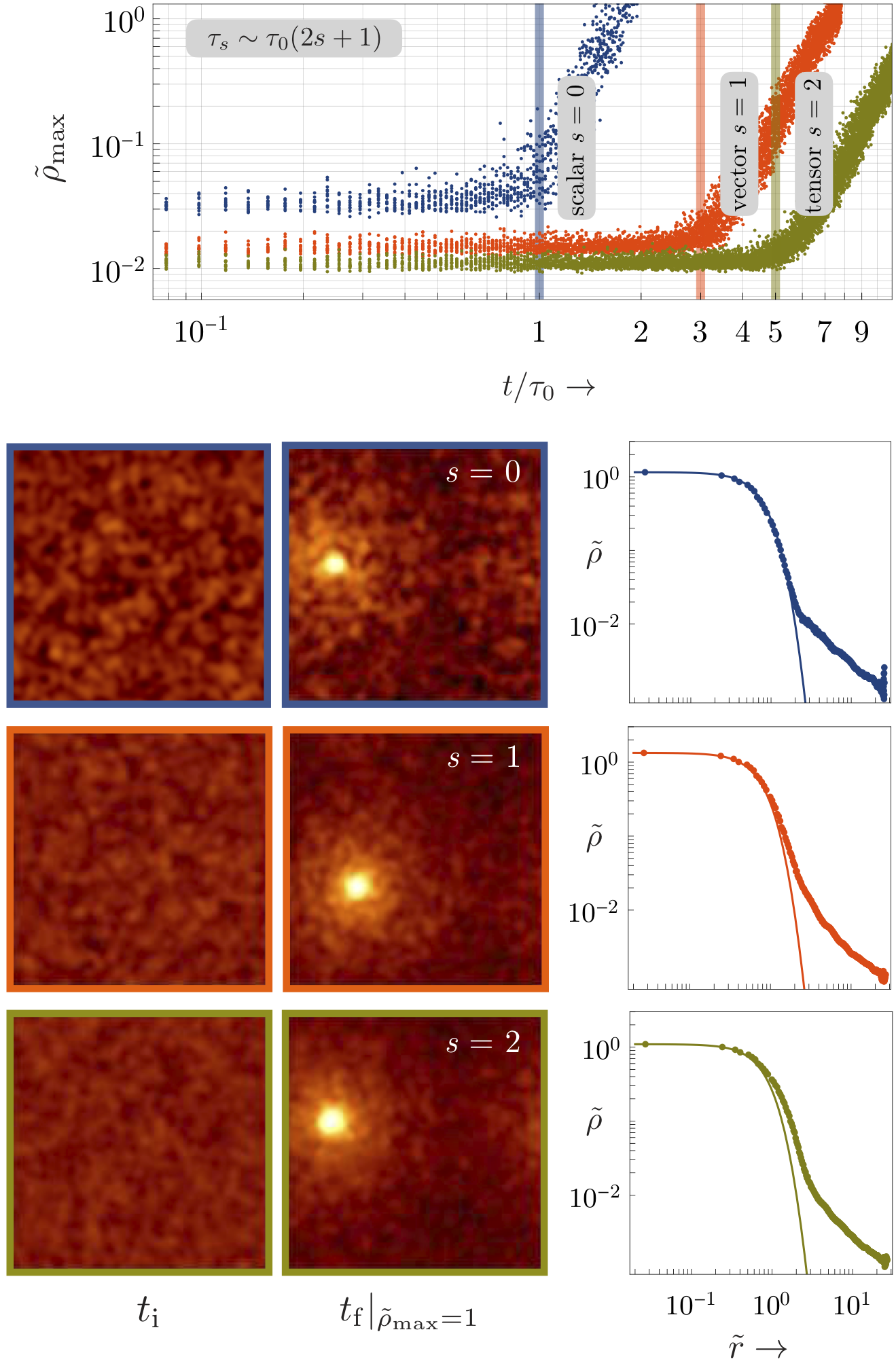}
    \caption{{\it Top panel}: Maximum density in the simulation volume as a function of time for scalar ($s=0$), vector ($s=1$) and tensor fields ($s=2$). The condensation time scales with the number of components of the field as $\tau_{(s)}\sim\tau_0\times N$, where $N=2s+1$. The simulated data includes $14$ simulations for $s=0,1,2$ each. For visual clarity, the output shown are significantly under-sampled compared to what is available from our simulations. {\it Lower panels}: In each row (corresponding to scalar, vector and tensor fields respectively), the first two panels show a projection of the mass density of the spin-$s$ field at initial and final times, while the third panel provides the radial profile of the mass density (solid line is the expected soliton profile) at the final time. Some simulation animations are available \href{https://mustafa-amin.com/home/multicomponent-dark-matter}{here}.}
    \label{fig:Plot2s+1}
\end{figure} 
\subsection{Equal mass, spin-$s$ case}\label{subsection:spin-s}

First we consider the case of a spin-$s$ field with $N=2s+1$ components, for which all the components have the same mass $m$. Assuming equipartition of mass density, i.e. $\bar{\rho}_a = \bar{\rho}/(2s+1)$ for all components where $\bar{\rho}$ is the total average mass density, alongwith equal velocity dispersion $\sigma$ for all components, the evolution equation (c.f~\eqref{eq:f0_evolveDE}) for any component becomes $\dot{g} = \tau^{-1}_{\rm gr}(2g^3 - \beta g)/(2s+1)$.\footnote{Here we have assumed that all the $\beta$ factors are same, owing to democratic initial conditions.} Notice that the only difference as compared to the scalar ($s=0$) case is that we have democratically populated all the components, giving rise to an overall $\bar{\rho}^2/(2s+1)^2$ factor, and a $2s+1$ factor owing to the summation over the $2s+1$ components (due to universality of gravity). The net result is a $1/(2s+1)$ factor in the rate of kinetic relaxation. Equivalently, the rate defined in \eqref{eq:Rate_initial} evaluates to $\Gamma_{(s)}=\Gamma_0/(2s+1)$. The time of condensate nucleation (within any component) is therefore estimated as
\begin{align}\label{eq:tau_s_prediction}
    \tau_{(s)} \sim \tau_{0}(2s+1)\,.
\end{align}
To verify the above prediction, we have performed $\sim 50$ simulations for $s=0,1$ and $2$ (corresponding to scalar, vector and tensor wavelike dark matter).\footnote{To verify the robustness of our scaling result $\tau \sim \tau_0\,N$, we also performed $\sim 10$ simulations for $N=2$ and $N=4$ cases.} We provide necessary details of the actual simulations in appendix~\ref{app:numerical_details}. Fig.~\ref{fig:Plot2s+1} shows our simulation results along with comparison with analytics. The densities are normalized by $(\sigma^2 m/\sqrt{G})^2$, and length scales by $1/(m\sigma)$. 

\begin{figure}[!b]
    \centering
    \includegraphics[width=0.5\textwidth]{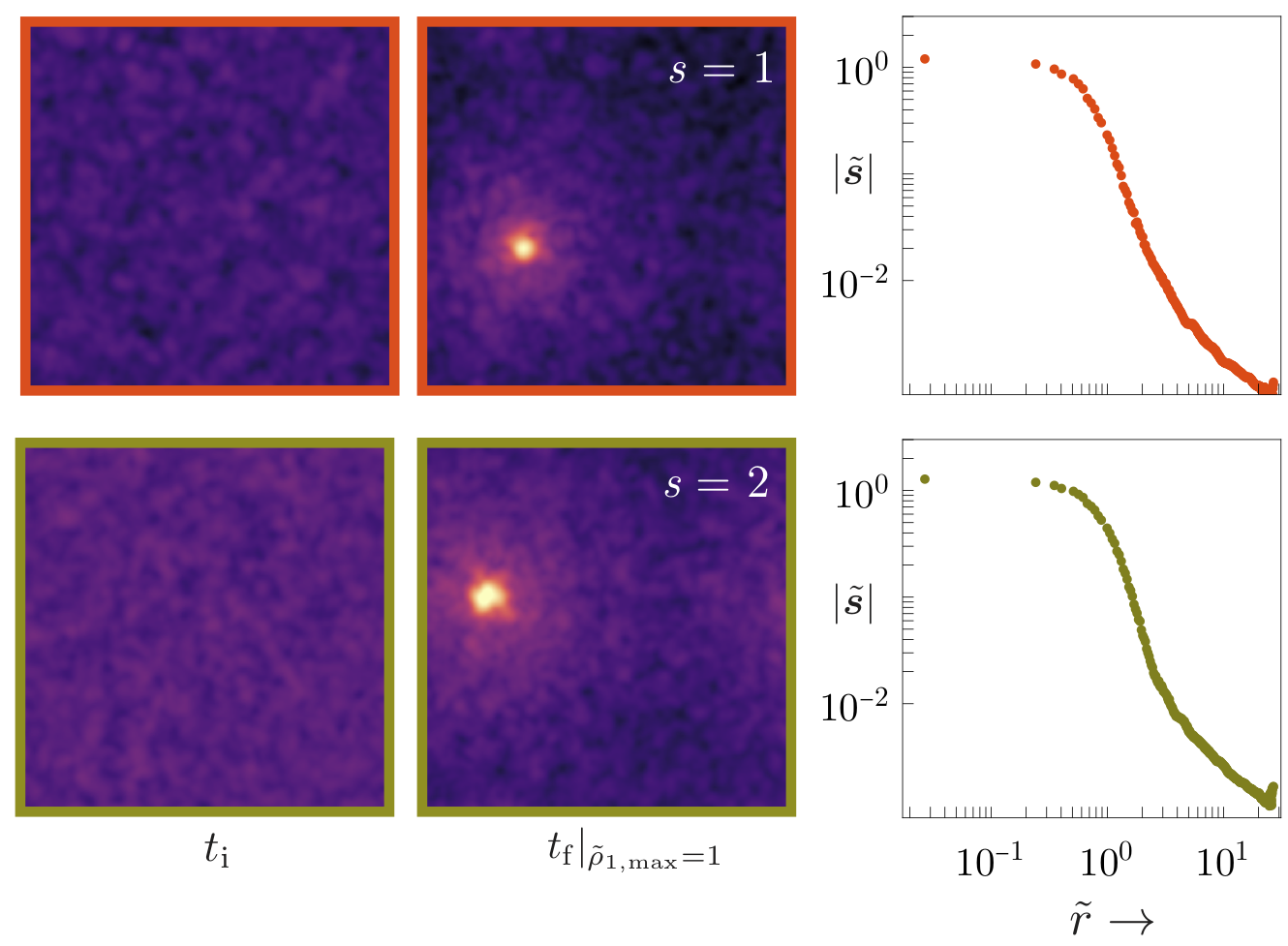}
    \caption{The simulation snapshots in the top and bottom row show the initial and final projections of the magnitude of the spin-density for vector and tensor cases respectively. The rightmost column show the radial profile of the magnitude of the spin density at the final time. Note that spin accumulates with the density (compare with bottom two rows of Fig.~\ref{fig:Plot2s+1}). Restoring factors of $\hbar$, the spin per boson in the simulation volume is $\mathcal{O}(10^{-2})\hbar$, whereas in the core it concentrates to $\mathcal{O}(1)\hbar$ . Unlike the magnitude of the radial spin density profile, spin in the core and in the simulation volume is obtained by vector summation of spin density at each location.}
    \label{fig:PlotSpin}
\end{figure} 

For simulations, we take the condensation time to be the time when there is a characteristic change in slope (on a log-log scale) of the maximum density in the simulation volume vs. time.  Note that the $\tau_0$ used to normalize the time axis in the top panel of Fig~\ref{fig:Plot2s+1} is extracted from simulations for the scalar case, chosen to highlight the scaling of the condensation time with the number of components.

The density in the box at initial times and after the soliton is reasonably well formed (we decided this based on a fixed density threshold $\tilde{\rho}_{\rm max}=1$) are also shown in the lower panels.  The soliton profile in total density shows good agreement with theoretical expectations \cite{Jain:2021pnk}.  We also kept track of densities in individual components of the fields. For the multicomponent cases (in particular the tensor one), not all components have the same shape of the density profile at the final snapshot shown. We see an increasing approach to similar profile shapes as time progresses and the agreement of the soliton profile with the theoretically expected one improves. Note the reduced interference effects (seen as less contrast in the colors, but the length scale of the patterns remains the same) in the initial conditions or in the patterns away from the soliton, as expected from~\cite{Amin:2022pzv}. The same phenomenon was also seen in~\cite{Gosenca:2023yjc}. The amplitude and length-scale of interference patterns has been used to constrain the mass of ultra-light dark matter \cite{Church:2019,2022arXiv220305750D,Powell:2023jns}.

\begin{figure}[!t]
    \centering
    \includegraphics[width=0.49\textwidth]{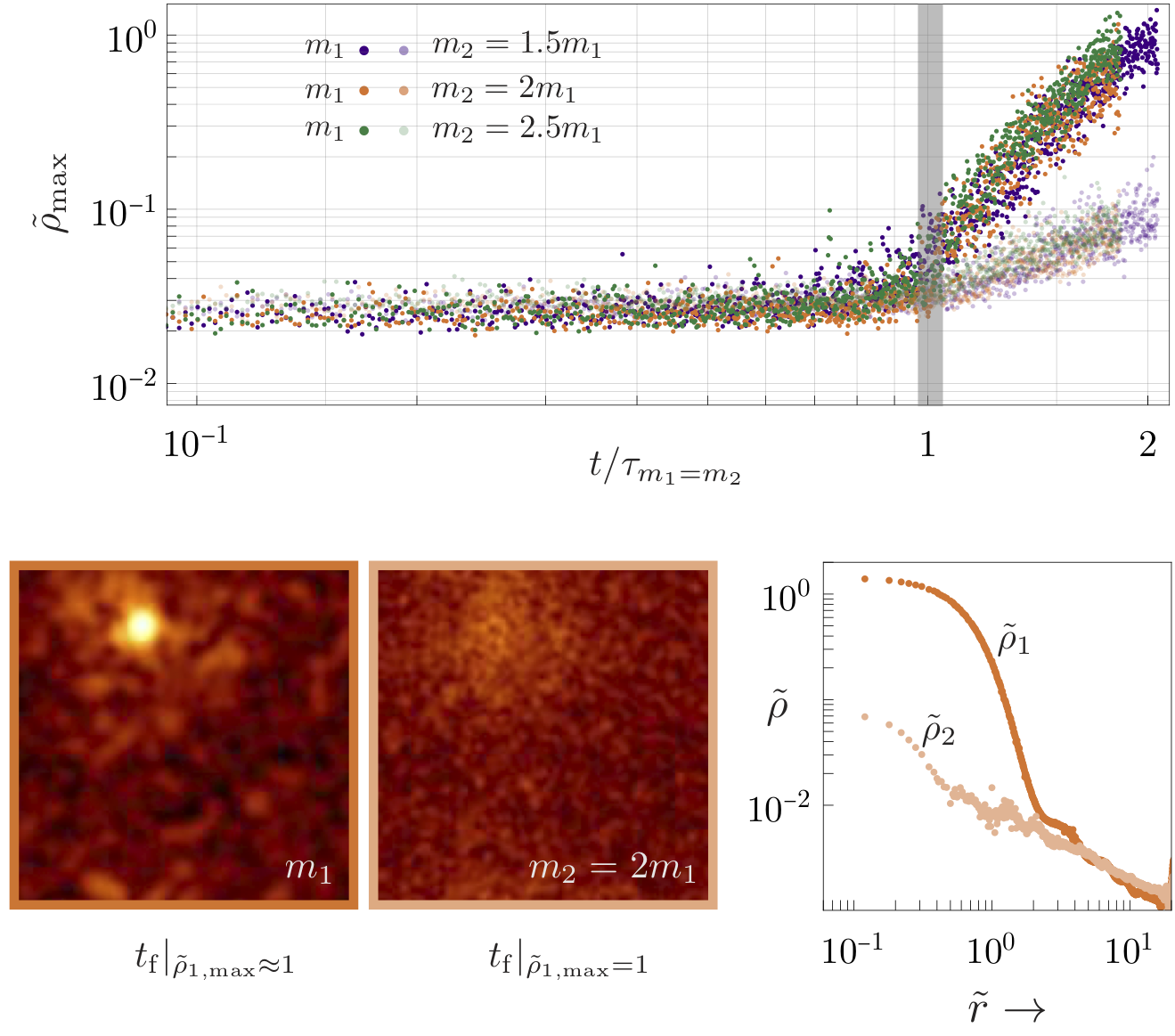}
    \caption{2-component simulations with equal mass density in each component, but different boson masses. The boson mass and mass density of the first component is held fixed. {\it Top panel}: Maximum density of each component of the field as a function of time in the simulation volume. Three simulations are shown, each with a different ratio of boson masses between the two components. Transparent version of each color corresponds to the heavier component. Note that there is no-significant dependence of the condensation time on the mass ratios considered here. Also note the slower accumulation rate at late times of the heavier component. {\it Bottom panel} : First two panels show final projected densities in the lighter and heavier components, whereas the third shows their radial profiles. The heavier component is accumulating around the condensed lighter one. Some simulation animations are available \href{https://mustafa-amin.com/home/multicomponent-dark-matter}{here}.}
    \label{fig:PlotEqRho}
\end{figure} 

Furthermore, we calculate the spin densities (see \cite{Jain:2021pnk,Amin:2022pzv}) of the condensates at final times in the respective simulation sets. We show spin density plots for the two cases in Fig.~\ref{fig:PlotSpin}. Note that the solitons that form have significant spin/boson at the end of the simulations.

\begin{figure}[!t]
    \centering
\includegraphics[width=0.485\textwidth]{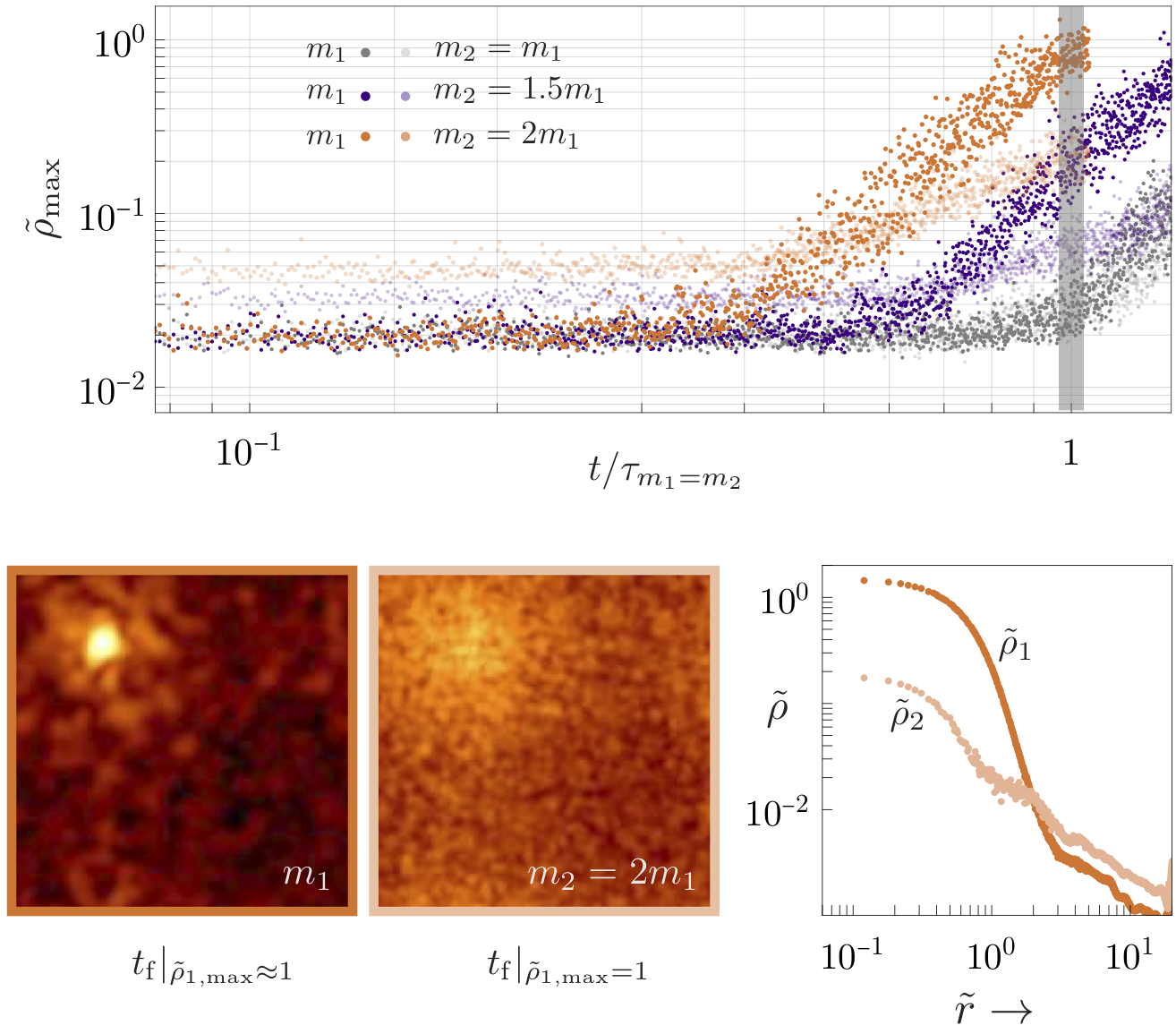}
    \caption{2-component simulations with equal number density in each component, but different boson masses. The boson mass and mass density of the first component is held fixed. {\it Top panel}: Two simulations are shown, each with a different ratio of boson masses between the two components. Transparent version of each color corresponds to the heavier component. In contrast with the equal mass density case, the condensation time decreases with increasing $m_2/m_1$.  {\it Bottom panel} : First two panels show final projected densities in the lighter and heavier components, whereas the third shows their radial profiles. Note that the difference in initial mass densities between the two components is still visible at large radii from the soliton's center.}
    \label{fig:PlotEqNum}
\end{figure} 

\subsection{Unequal masses : Multiple scalars}\label{subsection:unequal-masses}

\begin{figure}[!t]
    \centering
\includegraphics[width=0.485\textwidth]{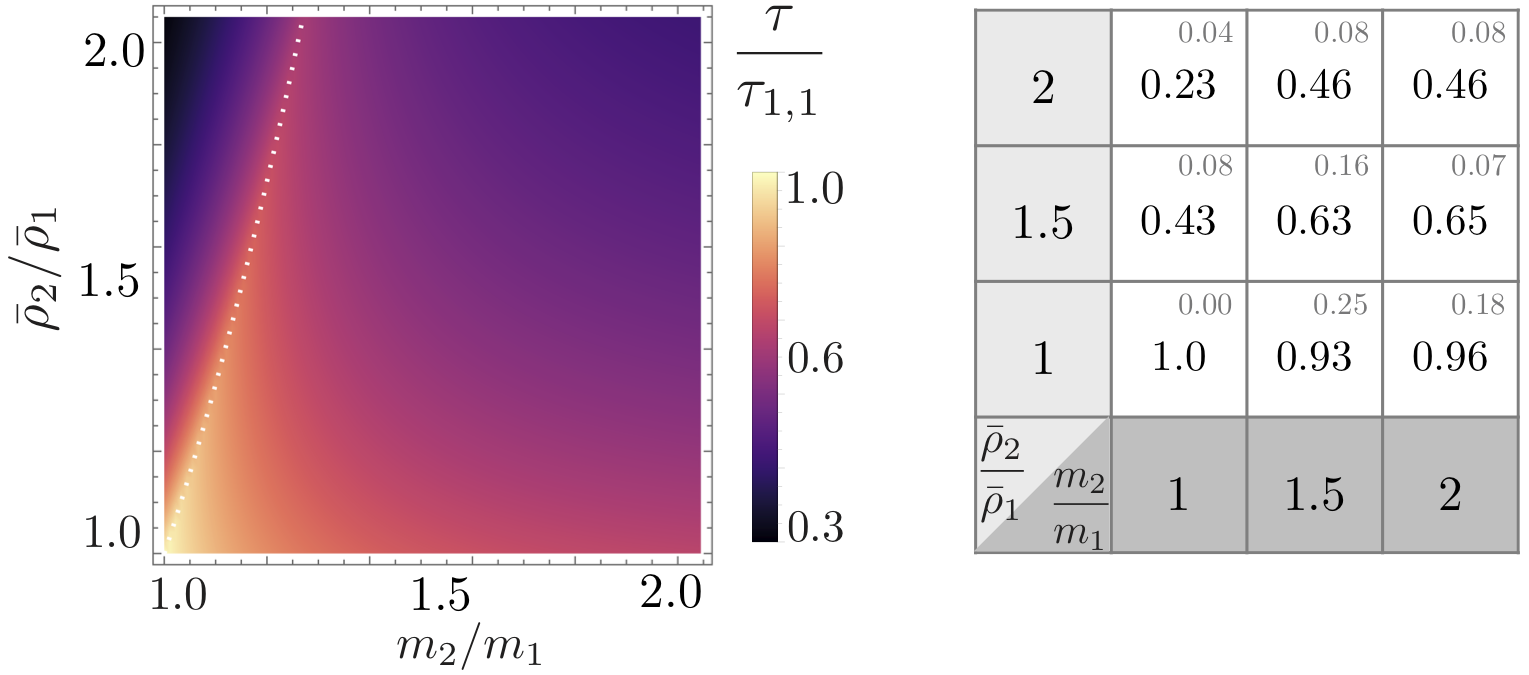}
    \caption{{\it Left Panel}: Analytical estimate for the time scale of emergence in two component systems with different boson masses and average mass densities, based on the initial kinetic relaxation rate Eq.~\eqref{eq:Rates_1_2}. To the right of the dotted line, component with boson mass $m_1$ condenses first, whereas to the left of the dotted line component with boson mass $m_2$ condenses first. Times are normalized by the equal density, equal boson mass case . We vary $m_2$ and $\bar{\rho}_2$, keeping $m_1$ and $\bar{\rho}_1$ fixed. {\it Right Table}: Condensation times (normalized by $\tau_{1,1}$ for each simulation set) extracted from numerical simulations. For each $\{\bar{\rho}_2/\bar{\rho}_1, m_2/m_1\}$ point, we have averaged over $5$ simulation runs with two different values of $\bar{\rho}_1$. The qualitative trends with density and mass ratios match the theoretical expectations. } 
    \label{fig:TheoryTable}
\end{figure}

Here, the different components are scalars with different masses. Focusing on the case of a two-component scalar dark matter but with the same characteristic velocity across each species (c.f. Eq~\eqref{eq:Gaussian_initialansatz} with $\sigma_a = \sigma$ for both $a = 1$ and $2$), we have from Eq.~\eqref{eq:f0_evolveDE}
\begin{align}\label{eq:g1_g2_eqns}
    \dot{g}_1 &= \frac{1}{\tau_{\rm gr}}\left(2 g_1(g_1 + yg_2) - \beta_{11} - \beta_{12}\frac{y^2}{x^3}\right)g_1\nonumber\\
    \dot{g}_2 &= \frac{1}{\tau_{\rm gr}}\left(\frac{2 y g_2(g_1 + yg_2)}{x^3} - \beta_{21} - \beta_{22}\frac{y^2}{x^3}\right)g_2\,,
\end{align}
where we have defined $m_1 = m$, $m_2 = x m$ and $\bar{\rho}_1 = \bar{\rho}$, $\bar{\rho}_2 = y\bar{\rho}$, and also recall that $\tau_{\rm gr} \equiv 2m^3\sigma^6/(\Lambda(4\pi G)^2\bar{\rho}^2)$. While in principle it is possible to estimate the $\beta$ parameters with the aid of a suite of simulations (under the assumption of them being more or less time independent), we don't perform this exercise in this paper. To get reasonable analytical insights, we rather work with the initial rate given by~\eqref{eq:Rate_initial}. With Gaussian initial conditions (i.e. $\beta = 1$), we get
\begin{align}\label{eq:Rates_1_2}
    \Gamma_1 &= \frac{\Gamma_{1,1}}{2}\left(2(1 + y) - 1 - \frac{y^2}{x^3}\right)\nonumber\\
    \Gamma_2 &= \frac{\Gamma_{1,1}}{2}\left(\frac{2y}{x^3}(1 + y) - 1 - \frac{y^2}{x^3}\right)\,,
\end{align}
where $\Gamma_{1,1}=\Gamma_{x=1,y=1}$, with the corresponding times for each component being $\tau_{a}/\tau_{1,1}\sim \Gamma_{1,1}/\Gamma_{a}$. Note that in the above, we have ignored Coulomb log factors which would appear when masses are unequal. We use this estimate (plotted in the left panel in Fig.~\ref{fig:TheoryTable}) to compare with a suite of simulations (right panel). In what is discussed below, we always keep the mass of the first component ($m_1=m$), and its density ($\bar{\rho}_1=\bar{\rho}$) fixed, while the same for the second component are varied using $x,y\ge 1$. 

Based on our simulations, we provide the behavior of maximum density vs. time, and the density snapshots and profiles of the nucleated solitons for equal mass density, and equal number density cases in Fig.~\ref{fig:PlotEqRho} and Fig.~\ref{fig:PlotEqNum} respectively. We provide a more statistical viewpoint of the condensation times in the table in Fig.~\ref{fig:TheoryTable}. In that table, we summarize our numerical findings for various values of $x$ and $y$. We carried out $5$ sets of $9$ (in total $45$) simulations to explore the dependence on mass densities and masses. We have provided both the average and the standard deviation resulting from different initial ``seeds" (different random phases) for each $x$ and $y$ value.

Some of the results are as follows:
\begin{enumerate}
\item For equal mass densities, $y =\bar{\rho}_2/\bar{\rho}_1= 1$, our estimate indicates that a condensate nucleates in the lighter field,  with its time of condensation eventually becoming independent of $x=m_2/m_1>1$, and approximately equal to $\tau_{1,1}$. This behaviour is seen in the top panel of Fig.~\ref{fig:PlotEqRho}, as well as the bottom row of the right panel in Fig.~\ref{fig:TheoryTable}.
\item For equal number densities between the two species, i.e. along the $y = x$ line, it is still the first species that forms a condensate, but the time scale of its nucleation decreasing as $\sim \tau_{1,1}/x$. This is again seen in the top panel of Fig.~\ref{fig:PlotEqNum} and the diagonal of the table in Fig.~\ref{fig:TheoryTable}. 
\item For equal masses ($x = 1$) but unequal mass densities ($y > 1$), we can see that now it is the second species within which a condensate nucleates first, with it's time of emergence eventually scaling as $\sim \tau_{1,1}/y^2$. We verify this trend in the first column of the table in Fig.~\ref{fig:TheoryTable}. 
\item Finally, Eq.~\eqref{eq:Rates_1_2} reveals a dividing curve $y = x^3$. To the left of this curve, the second component condenses faster and to the right, the first component condenses first (see left panel of Fig.~\ref{fig:TheoryTable}). For a constant $y$, to the left of $y=x^3$ the time of condensation of the heavier species is increasing with $x$ whereas that of the lighter one is decreasing, eventually crossing at $y=x^3$. To the right of this line the condensation time of the lighter species decreases but approaches a constant rapidly. We see this qualitatively in some of our simulations, with $y\sim x^3$ providing a rough guide for this change in behaviour.
\end{enumerate}

In the above analysis of the simulations, we kept the mass and density of the lighter component fixed and varied the mass and density of the heavier component. One can of course also keep the total density fixed (as we did in the spin-$s$ case). In this case we can parameterize $\rho_1=Y\bar{\rho}$ and $\rho_2=(1-Y)\bar{\rho}$, modifying the dividing line between which component condenses as $Y = 1/(1+x^3)$. The rest of the analysis is straightforward to carry out based on the initial rate equation.

\section{Summary}
\label{sec:summary}
In this paper we investigated kinetic relaxation in multicomponent Schr\"{o}dinger-Poisson (SP) system and nucleation of solitons. Starting with an $N$-component SP system with each component potentially having a different boson mass, we derived a Boltzmann equation in Fourier space for the occupation number function for each component (valid in the kinetic regime). Writing the Boltzmann equation in the Fokker-Planck form where the contributions from diffusion and friction terms become apparent, we discussed how the occupation number function for the condensing species evolve with time, with specific focus on its growth at vanishing momenta which is relevant for nucleation of condensates.

While we do not pursue numerical evolution of the coupled set of Boltzmann equation, we analyze its basic structure at small momenta (assuming quadratic functional dependence on the momenta), and provide a coupled set of ODEs for the evolution of the occupation number functions. This gives us a way to estimate the time scales of soliton nucleation. To analyze our estimates, we also performed full 3+1 dimensional simulations of the multicomponent SP system. For the purposes of simulations, we considered initial conditions where the initial field amplitudes (in Fourier space) had a Maxwell-Boltzmann like distribution with random phases. This was meant to mimic conditions inside halos. 

We have focused on two broad scenarios, results and comparisons with analytic estimates for which we outline below:
\begin{itemize}
    \item For the case of a massive spin-$s$ field, where the number of components $N=2s+1$, each component naturally has the same boson mass. Starting with democratic initial conditions, i.e. same average mass density in each component and equal velocity dispersion, we analytically estimate and numerically verify that the time-scale of condensation goes as $\tau_s \sim N\tau_0$ where  $\tau_0$ is the time of condensation for the single component case. Thus, under these initial conditions, solitons emerge later in higher-spin fields. Moreover, and as expected, we found that the spin density accumulates in the cores with spin magnitude per boson $\sim \mathcal{O}(1)$ even when starting with negligible initial spin magnitude/boson in the system.
    \item The second case we considered was a two-component system with different (but comparable) boson masses and average mass densities, and equal velocity dispersion. In general the mass density of each component, and corresponding boson mass can impact the condensation time-scale. Our analysis of condensation rate based on initial conditions, allowed us to estimate the time-scale for condensation in this general scenario and delineate regions in parameter space where one component condenses before the other. For a list of our results in this case, see the summary of results in Sec.~\ref{subsection:unequal-masses}.
\end{itemize}

We expect that our analysis of condensation rates and soliton formation in multicomponent SP systems in the kinetic regime should be useful for understanding the implications of such processes in cosmological and astrophysical settings. The formation rates depend mainly on ``local" conditions such as the density and velocity dispersion, however these in turn can be affected by the dark matter formation mechanism, including features in the density power spectrum at small scales (see, for example,~\cite{Hogan:1988mp,Graham:2015rva,Agrawal:2018vin,Co:2018lka,Dror:2018pdh,Bastero-Gil:2018uel,Irsic:2019iff,Co:2021rhi,Redi:2022llj,Adshead:2023qiw}). Such features in the power spectrum are generic in most post-inflationary production scenarios of light dark matter~\cite{Amin:2022nlh}, and present in many inflationary ones as well (see, for example,~\cite{Graham:2015rva}). 

We have focused on condensation via gravitational interactions alone in this paper. A natural generalization is to include non-gravitational self-interactions, especially in the case of a single spin-$1$ field which admits attractive self-interactions in the Higgs phase, or non-Abelian spin-$1$ set of fields which also admit repulsive self-interactions apart from the Higgs induced attractive ones~\cite{Zhang:2021xxa,Jain:2022kwq}. 
Related work of kinetic Bose condensation in a single scalar field was done in~\cite{Semikoz:1994zp,Kirkpatrick:2020fwd,Chen:2021oot}. In an upcoming publication, we will investigate the impact of such self-interactions on kinetic condensation time scales in the multicomponent case. \\ 

\noindent{\it Note added after first submission to the arXiv:} We note that another paper~\cite{Chen:2023bqy} appeared on the arXiv concurrently with this one, exploring the kinetic condensation in non-relativistic {\it vector} DM. Their numerical results agree with our general results where there is overlap: For their uncorrelated case, see our section~\ref{subsection:spin-s}). For their ``correlated" case (which is equivalent to lesser number of uncorrelated/statistically independent components with different average number densities), see end of our section~\ref{subsection:unequal-masses}.
\section*{Acknowledgements}
We thank P. Mocz for sharing his numerical code for condensation in a scalar SP system, and Dorian Amaral and Andrew Long for helpful discussions. MJ acknowledges useful discussions with Jiajun Chen and Xiaolong Du, after the first submission of this work to the arXiv. MA and MJ are partially supported by a NASA grant 80NSSC20K0518. JT and WW acknowledge summer support from the Department of Physics and Astronomy at Rice University. 

\bibliography{reference}
\appendix

\section{Wave kinetic equation for arbitrary $2 \rightarrow 2$ multicomponent-wave interactions}\label{app:wave_kinetic_derivation}

In this appendix we derive the wave kinetic equation for multicomponent \schr (non-relativistic) systems with arbitrary $2$-body scattering interactions. See~\cite{Zakharov1992KolmogorovSO} for a discussion for a single species of waves. In our derivation, we work with a finite box of volume $V$ and hence a discrete set of ${\bm k}$ values, and only towards the end of the calculation shall take the continuous limit. Using the Fourier decomposition $\psi^a({\bm x},t) = V^{-1/2}\sum_{\bm k}e^{-i{\bm k}\cdot{\bm x}}\,\Psi^a_{\bm k}(t)$, the \schr equation takes the following general form in ${\bm k}$ space
\begin{align}\label{eq:Schreqn_kspace}
    i\dot{\Psi}^a_{\bm k} &= E^a_k\Psi^a_{\bm k}\nonumber\\
    & + \frac{1}{V}\sum_{\bm p, \bm q, \bm \ell}\delta_{\bm k + \bm p - \bm q - \bm \ell}\sum_{b,c,d}\Bigl\{\mathcal{T}^{a,b,c,d}_{\bm k, \bm p, \bm q, \bm \ell}\,\Psi^{b\,\ast}_{\bm p}\Psi^c_{\bm q}\Psi^d_{\bm \ell}\Bigr\}\,,
\end{align}
where $E^a_k = k^2/2m_a$ is the free wave dispersion relation (for every species ``$a$"). The quantity $\mathcal{T}^{a,b,c,d}_{\bm k, \bm p, \bm q, \bm \ell}$ is the form factor (of mass dimension $-2$) that governs the structure of self-interactions, and has the following two properties
\begin{align}\label{eq:T_props}
    \mathcal{T}^{a,b,c,d\,\ast}_{\bm k, \bm p, \bm q, \bm \ell} &= \mathcal{T}^{d,c,b,a}_{\bm \ell, \bm q, \bm p, \bm k}\nonumber\\
    \mathcal{T}^{a,b,c,d}_{\bm k, \bm p, \bm q, \bm \ell} &= \mathcal{T}^{b,a,d,c}_{\bm p, \bm k, \bm \ell, \bm q}\,.
\end{align}
Both of these can be obtained by noting that the 
interaction Hamiltonian has the structure 
\begin{align}\label{eq:H_int}
    H_{\rm int} = \frac{1}{2V}\sum_{a,b,c,d}\sum_{\bm k, \bm p, \bm q, \bm \ell}\delta_{\bm k + \bm p - \bm q - \bm \ell}\,\mathcal{T}^{a,b,c,d}_{\bm k, \bm p, \bm q, \bm \ell}\,\Psi^{a\,\ast}_{\bm k}\Psi^{b\,\ast}_{\bm p}\Psi^{c}_{\bm q}\Psi^{d}_{\bm \ell}\,.
\end{align}
The realness of the Hamiltonian enforces the first property, whereas the symmetry under interchange of both incoming ($a$, $b$) and outgoing ($c$, $d$) species, carrying momenta ($\bm k$, $\bm p$) and ($\bm q$, $\bm \ell$) respectively, enforces the second property.
 
Breaking up the Fourier field $\Psi$ into an occupation number function $f$ and a phase function $\theta$, i.e. $\Psi^a_{\bm k} = \sqrt{f^a_{\bm k}}\,e^{-i\theta^a_{\bm k}}$, Eq.~\eqref{eq:Schreqn_kspace} gives
\begin{align}\label{eq:eqn_occupation_n_phase}
    \dot{f}^a_{\bm k} =&\, \frac{2}{V}\sum_{\bm p, \bm q, \bm \ell}\delta_{\bm k + \bm p - \bm q - \bm \ell}\sum_{b,c,d}\Im \Bigl[\mathcal{T}^{a,b,c,d}_{\bm k, \bm p, \bm q, \bm \ell}\,\mathcal{A}^{a,b,c,d}_{\bm k, \bm p, \bm q, \bm \ell}\Bigr]\,,\\
    f^a_{\bm k}\dot{\theta}^a_{\bm k} =&\, \frac{1}{V}\sum_{\bm p, \bm q, \bm \ell}\delta_{\bm k + \bm p - \bm q - \bm \ell}\sum_{b,c,d}\Re \Bigl[\mathcal{T}^{a,b,c,d}_{\bm k, \bm p, \bm q, \bm \ell}\,\mathcal{A}^{a,b,c,d}_{\bm k, \bm p, \bm q, \bm \ell}\Bigr]\nonumber\\
    &\, + f^a_{\bm k}E^a_k\,,\nonumber
\end{align}
where
\begin{align}\label{eq:A_define}
    \mathcal{A}^{a,b,c,d}_{\bm k, \bm p, \bm q, \bm \ell} = \Psi^{a\,\ast}_{\bm k}\Psi^{b\,\ast}_{\bm p}\Psi^{c}_{\bm q}\Psi^{d}_{\bm \ell} = \sqrt{f^a_{\bm k}f^b_{\bm p}f^c_{\bm q}f^d_{\bm \ell}}\,e^{i( \theta^a_{\bm k} + \theta^b_{\bm p} - \theta^c_{\bm q} - \theta^d_{\bm \ell})}\,.
\end{align}
Note that $f^a_{\bm k}(t)$ is nothing but the Fourier transform of the two-point correlation function $\int\mathrm{d}{\bm y}\,\psi^{a}(\bm{x},t)^{\ast}\psi^{a}(\bm{x} + \bm{y},t)$\,.

Now we wish to obtain an equation for the occupation number function alone. We will work in the small interaction regime where the typical time scale of oscillation of a single free ``$a$" type wave, $\tau^a_{\rm free} = 2m_a/k^2$, is very small as compared to the time scales associated with self-interactions. More formally, we impose $|\tau^a_{\rm free}\,d\{f^a,\theta^a\}^{n+1}/dt^{n+1}| \ll |d\{f^a,\theta^a\}^n/dt^n|$ consistently for all $n \geq 0$, for all species. Here $n$ would dictate the order in our perturbation scheme. 

Small interactions further dictate that since the free wave dispersion relation holds at leading order, phases $\theta^a_{\bm k}$ randomize over time irrespective of whether they were initially correlated or not. Hence for time scales much longer than $\tau_{\rm free}$, it is sufficient to work within the random phase approximation regime where phases are taken to be uncorrelated. Representing integration over phases by bra-kets, at leading ($n=0$) order we require $\langle e^{i(\theta^a_{\bm k} + \theta^b_{\bm p} - \theta^c_{\bm q} - \theta^d_{\bm \ell})}\rangle = \delta^{ac}\delta_{\bm k,\bm q}\delta^{bd}\delta_{\bm p,\bm \ell} + \delta^{ad}\delta_{\bm k,\bm \ell}\delta^{bc}\delta_{\bm p,\bm q}$ to obtain $\langle\mathcal{A}^{a,b,c,d}_{\bm k, \bm p, \bm q, \bm \ell}\rangle$ in Eq.~\eqref{eq:A_define}, which ultimately fetches $\dot{f}^a = 0$. Therefore, we need to go to the next order ($n=1$) to capture effects due to interactions. This requires setting the time derivative of $\mathcal{A}$ in~\eqref{eq:A_define}, after using the equations of motion~\eqref{eq:eqn_occupation_n_phase} and then integrating out the phases, to zero. This exercise yields
\begin{align}
    &\langle\mathcal{A}^{a,b,c,d}_{\bm k, \bm p, \bm q, \bm \ell}\rangle = \frac{-1}{\Delta E + i\epsilon}\times\nonumber\\
    &\; \Biggl[\frac{1}{f^a_{\bm k}}\frac{1}{V}\sum_{\bm k_1, \bm k_2, \bm k_3}\delta_{\bm k + \bm k_1 - \bm k_2 - \bm k_3}\sum_{a_1,a_2,a_3}\Bigl\{\mathcal{T}^{a,a_1,a_2,a_3\,\ast}_{\bm k, \bm k_1, \bm k_2, \bm k_3}\,\times\nonumber\\
    &\qquad\qquad\qquad\qquad\qquad\qquad\langle\mathcal{A}^{a,a_1,a_2,a_3\,\ast}_{\bm k, \bm k_1, \bm k_2, \bm k_3}\mathcal{A}^{a,b,c,d}_{\bm k, \bm p, \bm q, \bm \ell}\rangle\Bigr\}\nonumber\\
    &\; + \frac{1}{f^b_{\bm p}}\frac{1}{V}\sum_{\bm k_1, \bm k_2, \bm k_3}\delta_{\bm p + \bm k_1 - \bm k_2 - \bm k_3}\sum_{a_1,a_2,a_3}\Bigl\{\mathcal{T}^{b,a_1,a_2,a_3\,\ast}_{\bm p, \bm k_1, \bm k_2, \bm k_3}\,\times\nonumber\\
    &\qquad\qquad\qquad\qquad\qquad\qquad\langle\mathcal{A}^{b,a_1,a_2,a_3\,\ast}_{\bm p, \bm k_1, \bm k_2, \bm k_3}\mathcal{A}^{a,b,c,d}_{\bm k, \bm p, \bm q, \bm \ell}\rangle\Bigr\}\nonumber\\
    &\; - \frac{1}{f^c_{\bm q}}\frac{1}{V}\sum_{\bm k_1, \bm k_2, \bm k_3}\delta_{\bm q + \bm k_1 - \bm k_2 - \bm k_3}\sum_{a_1,a_2,a_3}\Bigl\{\mathcal{T}^{c,a_1,a_2,a_3}_{\bm q, \bm k_1, \bm k_2, \bm k_3}\,\times\nonumber\\
    &\qquad\qquad\qquad\qquad\qquad\qquad\langle\mathcal{A}^{c,a_1,a_2,a_3}_{\bm q, \bm k_1, \bm k_2, \bm k_3}\mathcal{A}^{a,b,c,d}_{\bm k, \bm p, \bm q, \bm \ell}\rangle\Bigr\}\nonumber\\
    &\; - \frac{1}{f^d_{\bm \ell}}\frac{1}{V}\sum_{\bm k_1, \bm k_2, \bm k_3}\delta_{\bm \ell + \bm k_1 - \bm k_2 - \bm k_3}\sum_{a_1,a_2,a_3}\Bigl\{\mathcal{T}^{d,a_1,a_2,a_3}_{\bm \ell, \bm k_1, \bm k_2, \bm k_3}\,\times\nonumber\\
    &\qquad\qquad\qquad\qquad\qquad\qquad\langle\mathcal{A}^{d,a_1,a_2,a_3}_{\bm \ell, \bm k_1, \bm k_2, \bm k_3}\mathcal{A}^{a,b,c,d}_{\bm k, \bm p, \bm q, \bm \ell}\rangle\Bigr\}
    \Biggr]\,,
\end{align}
where $\Delta E = E^a_k + E^b_p - E^c_q - E^d_\ell$, and we have added a $+i\epsilon$ to regulate the divergence when $\Delta E = 0$. The sign can be obtained by requiring that free waves die out in the infinite past. Another equivalent way is to consider adiabatic turning on of the interactions as time goes on. Using the definition~\eqref{eq:A_define} and the following identity (due to uncorrelated statistics owing to random phase approximation)
\begin{align}\label{eq:six_point}
    & \langle e^{i(\theta^b_{\bm p}-\theta^c_{\bm q}-\theta^d_{\bm \ell}-\theta^{a_1}_{\bm k_1}+\theta^{a_2}_{\bm k_2}+\theta^{a_3}_{\bm k_3})}\rangle =\nonumber\\
    &\quad \delta^{b,c}\delta_{\bm p, \bm q}(\delta^{d,a_3}\delta_{\bm \ell, \bm k_3}\,\delta^{a_1,a_2}\delta_{\bm k_1, \bm k_2} + \delta_{\bm \ell, \bm k_2}\delta^{d,a_2}\,\delta_{\bm k_1, \bm k_3}\delta^{a_1,a_3})\nonumber\\
    & + \delta^{b,d}\delta_{\bm p, \bm \ell}(\delta^{c,a_2}\delta_{\bm q, \bm k_2}\,\delta^{a_1,a_3}\delta_{\bm k_1, \bm k_3} + \delta_{\bm q, \bm k_3}\delta^{c,a_3}\,\delta_{\bm k_1, \bm k_2}\delta^{a_1,a_2})\nonumber\\
    & + \delta^{b,a_1}\delta_{\bm p, \bm k_1}(\delta^{c,a_2}\delta_{\bm q, \bm k_2}\,\delta^{d,a_3}\delta_{\bm \ell, \bm k_3} + \delta_{\bm q, \bm k_3}\delta^{c,a_3}\,\delta_{\bm \ell, \bm k_2}\delta^{d,a_2})\,,
\end{align}
we get
\begin{align}
    &\langle\mathcal{A}^{a,b,c,d}_{\bm k, \bm p, \bm q, \bm \ell}\rangle = \frac{-\delta_{\bm k + \bm p - \bm q - \bm \ell}}{V(\Delta E + i\epsilon)}\;\Biggl[\left(\mathcal{T}^{a,b,c,d\,\ast}_{\bm k, \bm p, \bm q, \bm \ell} + \mathcal{T}^{a,b,d,c\,\ast}_{\bm k, \bm p, \bm \ell, \bm q}\right)f^b_{\bm p}f^c_{\bm q}f^d_{\bm \ell}\nonumber\\
    &\qquad\qquad\qquad\qquad + \left(\mathcal{T}^{b,a,c,d\,\ast}_{\bm p, \bm k, \bm q, \bm \ell} + \mathcal{T}^{b,a,d,c\,\ast}_{\bm p, \bm k, \bm \ell, \bm q}\right)f^a_{\bm k}f^c_{\bm q}f^d_{\bm \ell}\nonumber\\
    &\qquad\qquad\qquad\qquad - \left(\mathcal{T}^{c,d,b,a}_{\bm q, \bm \ell, \bm p, \bm k} + \mathcal{T}^{c,d,a,b}_{\bm q, \bm \ell, \bm k, \bm p}\right)f^a_{\bm k}f^b_{\bm p}f^d_{\bm \ell}\nonumber\\
    &\qquad\qquad\qquad\qquad - \left(\mathcal{T}^{d,c,b,a}_{\bm \ell, \bm q, \bm p, \bm k} + \mathcal{T}^{d,c,a,b}_{\bm \ell, \bm q, \bm k, \bm p}\right)f^a_{\bm k}f^b_{\bm p}f^c_{\bm q}\Biggr]\,.
\end{align}
We note that only the last line in the identity~\eqref{eq:six_point} ends up contributing (on account of the general properties~\eqref{eq:T_props} of the form factor $\mathcal{T}$). Extraction of the imaginary part of the above expression (needed for $\dot{f}$ as in~\eqref{eq:eqn_occupation_n_phase}), may be most easily done by going to the continuous regime. With $\sum_{\bm k} \rightarrow V(2\pi)^{-3}\int\mathrm{d}{\bm k}$ and $\delta_{\bm k, \bm p} \rightarrow V^{-1}(2\pi)^3\delta^{3}(\bm k - \bm p)$, along with using $\Im{(x + i\epsilon)^{-1}} = -\pi\delta(x)$ to regulate the divergence, we get\footnote{Here we also discarded the redundant momentum conservation Kronecker delta $\delta_{\bm k + \bm p - \bm q - \bm \ell}$.}
\begin{widetext}
    \begin{align}\label{eq:fdot_multicomponent_general}
    \dot{f}^a_{\bm k} &= \sum_{b,c,d}\int\frac{\mathrm{d}{\bm p}}{(2\pi)^3}\,\mathrm{d}\sigma_{\bm k_a + \bm p_b \rightarrow \bm q_c + \bm \ell_d}\,|{\bm v}_a-\tilde{\bm v}_b|\,\Biggl[(f^a_{\bm k} + f^b_{\bm p})f^c_{\bm q}f^d_{\bm \ell} - (f^c_{\bm q} + f^d_{\bm \ell})f^a_{\bm k}f^b_{\bm p}\Biggr]\,\quad {\rm where},\nonumber\\
    \mathrm{d}\sigma_{\bm k_a + \bm p_b \rightarrow \bm q_c + \bm \ell_d} &= \frac{\mathrm{d}{\bm q}}{(2\pi)^3}\frac{\mathrm{d}{\bm \ell}}{(2\pi)^3}\frac{1}{|{\bm v}_a-\tilde{\bm v}_b|}\mathcal{T}^{a,b,c,d}_{\bm k, \bm p, \bm q, \bm \ell}\Bigl(\mathcal{T}^{a,b,c,d}_{\bm k, \bm p, \bm q, \bm \ell} + \mathcal{T}^{a,b,d,c}_{\bm k, \bm p, \bm \ell, \bm q}\Bigr)^{\ast}\,\times\nonumber\\
    &\qquad\qquad\qquad\qquad\qquad\qquad\qquad\qquad\qquad (2\pi)^4\,\delta^{(3)}\left(\bm k + \bm p - \bm q - \bm \ell\right)\delta\left(E^a_k + E^b_p - E^c_q - E^d_\ell\right)\,.
    \end{align}
\end{widetext}
Here we have defined incoming ``velocities" ${\bm v}_a = {\bm k}/m_a$ and $\tilde{v}_{b} = {\bm p}/m_b$, and also used~\eqref{eq:T_props} to rewrite form factors to give a compact structure in terms of the differential cross section. Eq.~\eqref{eq:fdot_multicomponent_general} is the master wave kinetic equation for \textit{any} multicomponent \schr system with $2$-body self interactions dictated by the form factor $\mathcal{T}$ (c.f. Eq.~\eqref{eq:H_int}). For our purposes in this paper, we only focus on gravitational interactions, for which 
\begin{equation}
    \mathcal{T}^{a,b,c,d}_{\bm k, \bm p, \bm q, \bm \ell} = -(4\pi G)m_a m_b\,\delta_{bc}\delta_{da}|{\bm k}-{\bm \ell}|^{-2}.
\end{equation}
Using this in Eq.~\eqref{eq:fdot_multicomponent_general} gives eq.~\eqref{eq:Boltzmann_gravity} presented in the main text.

\section{Collision integral in the Eikonal approximation}\label{sec:app_eikonal_approx}

In order to get analytical insights, we approximate the collision term in the Boltzmann equation, in an eikonal approximation. Starting from the wave kinetic equation~\eqref{eq:Boltzmann_gravity}, we first massage it into a more digestible form by working with relative velocities. For this purpose, let us redefine $\frac{{\bm p}}{m_b} = \frac{{\bm k}}{m_a} - {\bm u}$ and $\frac{{\bm q}}{m_b} = \frac{{\bm \ell}}{m_a} - {\bm u}'$, where ${\bm u}$ and ${\bm u}'$ are the relative velocity vectors before and after the interaction process ${\bm k}_a + {\bm p}_b \rightarrow {\bm \ell}_a + {\bm q}_b$. It also becomes apparent that the magnitude of the relative velocity doesn't change during the process (a general property of two body elastic collisions), in practice enforced by energy conservation. Integration over ${\bm \ell}$ and $|{\bm u}'|$ yields
\begin{widetext}
    \begin{align}\label{eq:Boltzmann_gravity_com}
    &\frac{\partial f^a_{\bm v_a}}{\partial t} = \sum_{b}m_b^3\int\frac{\mathrm{d}\Omega_n}{4\pi}\frac{\mathrm{d}u\,u^2}{2\pi^2}\mathrm{d}\sigma\,u\Biggl[(f^a_{\bm v_a} + f^b_{\tilde{{\bm v}}_b})f^a_{\bm v_a + {\bm w}/m_a}f^b_{\tilde{{\bm v}}_b - {\bm w}/m_b} - (f^a_{\bm v_a + {\bm w}/m_a} + f^b_{\tilde{{\bm v}}_b - {\bm w}/m_b})f^a_{\bm v_a}f^b_{\tilde{{\bm v}}_b}\Biggr]\,,
    \nonumber\\
    &\qquad \qquad {\rm where} \quad \mathrm{d}\sigma = \frac{\mathrm{d}\Omega_{n'}}{4\pi^2}\,\frac{(4\pi m_am_bG)^2}{\mu^2 u^4|\hat{\bm n}' - \hat{\bm n}|^2}\Biggl(\frac{1}{|\hat{\bm n}' - \hat{\bm n}|^2} + \frac{\delta_{ab}}{|\hat{\bm n}' + \hat{\bm n}|^2}\Biggr)\,.
\end{align}
\end{widetext}
Here ${\bm v}_a = {\bm k}/m_a$ is the velocity vector for incoming wave type $a$, $\tilde{{\bm v}}_b = {\bm v}_a - u\hat{\bm n}$ is the velocity vector for incoming wave type $b$, and $\hat{\bm n}$, $\hat{\bm n}'$ are unit vectors in the direction of ${\bm u}$ and ${\bm u}'$ respectively (with $\Omega_n$, $\Omega_{n'}$ being the associated angular integral measures). Furthermore with $\mu = m_am_b/(m_a+m_b)$ as the reduced mass, ${\bm w} = \mu u (\hat{\bm n}'-\hat{\bm n})$ is the change in the momentum of species $a$ on account of interaction (before and after). Finally, for convenience, we have re-scaled the occupation number functions by the respective masses (rendering them functions of velocities).

The above simplification is a reflection of the fact that in any elastic collision, only the direction of the relative velocity changes. 
To progress further, since the differential cross section is dominated by small values of $|\hat{\bm n}'-\hat{\bm n}|$, we can expand the occupation number functions containing ${\bm w}$, around $\bm 0$. Physically this means that the change in relative velocities of two interacting waves is expected to be small for most of the interactions. The expansion is
\begin{align}
    f^a_{\bm v_a + {\bm w}/m_a} &= f^a_{\bm v_a} + \frac{1}{m_a} {\bm w}\cdot\nabla_{\bm v_a}f^a_{\bm v_a}\nonumber\\
    &\qquad + \frac{1}{2m_a^2}\,w^i\,w^j\,\nabla_{v^i_a}\nabla_{v^j_a}f^a_{\bm v_a} + ...\nonumber\\
    f^b_{\tilde{{\bm v}}_b - {\bm w}/m_b} &= f^b_{\tilde{{\bm v}}_b} - \frac{1}{m_b}{\bm w}\cdot\nabla_{\tilde{{\bm v}}_b}f^b_{\tilde{{\bm v}}_b}\nonumber\\
    &\qquad + \frac{1}{2m_b^2}\,w^i\,w^j\,\nabla_{\tilde{v}^i_b}\nabla_{\tilde{v}^j_b}f^b_{\tilde{{\bm v}}_b} + ...
\end{align}
After some algebra, we get the following result (up to quadratic order in ${\bm w} = \mu u(\hat{\bm n}'-\hat{\bm n})$):
\begin{align}
        &\frac{\partial f^a_{{\bm v}_a}}{\partial t} = \sum_{b}m_b^3\int\frac{\mathrm{d}\Omega_n}{4\pi}\frac{\mathrm{d}u\,u^2}{2\pi^2}\,\mathrm{d}\sigma\,u\;\times\nonumber\\
        &\qquad\quad\Biggl[\frac{1}{m_a}f^b_{\tilde{{\bm v}}_b}f^b_{\tilde{{\bm v}}_b}{\bm w}\cdot\nabla_{{\bm v}_a}f^a_{{\bm v}_a} - \frac{1}{m_b}f^a_{{\bm v}_a}f^a_{{\bm v}_a}{\bm w}\cdot\nabla_{\tilde{{\bm v}}_b}f^b_{\tilde{{\bm v}}_b}\nonumber\\
        &\qquad\quad - \frac{1}{m_am_b}(f^a_{{\bm v}_a} + f^b_{\tilde{{\bm v}}_b})({\bm w}\cdot\nabla_{{\bm v}_a}f^a_{{\bm v}_a})({\bm w}\cdot\nabla_{\tilde{{\bm v}}_b}f^b_{\tilde{{\bm v}}_b})\nonumber\\
        &\qquad\quad + \frac{1}{2m_a^2}f^b_{\tilde{{\bm v}}_b}f^b_{\tilde{{\bm v}}_b} w^iw^j\nabla_{v^i_a}\nabla_{v^j_a} f^a_{{\bm v}_a}\nonumber\\
        &\qquad\quad + \frac{1}{2m_b^2}f^a_{{\bm v}_a}f^a_{{\bm v}_a} w^iw^j\nabla_{\tilde{v}^i_b}\nabla_{\tilde{v}^j_b}f^b_{\tilde{{\bm v}}_b}\Biggr]\,.
\end{align}
Now to evaluate the $\Omega_{n'}$ integral, we can easily set $\hat{\bm n}'-\hat{\bm n} = (\sin\theta\cos\phi, \sin\theta\sin\phi, \cos\theta-1)$ with $\mathrm{d}\Omega_{n'} = \mathrm{d}\phi\,\mathrm{d}(\cos\theta)$, and $\theta$ integral restricted to small values (on account of our approximation $|\hat{\bm n}'-\hat{\bm n}| \ll 1$). Then with $|\hat{\bm n}'-\hat{\bm n}| = 2\sin(\theta/2)$, the two integrals that are relevant, are
\begin{align}
    &\frac{1}{2\pi^2}\int\mathrm{d}(\cos\theta)\int^{2\pi}_0\mathrm{d}\phi\,\frac{(\hat{n}'-\hat{n})_i}{16\sin^4(\theta/2)} = \nonumber\\
    &\qquad\qquad\qquad\qquad\qquad - \frac{\hat{n}_i}{2\pi}\log\left(\frac{\sin(\theta_{\rm max}/2)}{\sin(\theta_{\rm min}/2)}\right)\,,\nonumber\\
    &\frac{1}{2\pi^2}\int\mathrm{d}(\cos\theta)\int^{2\pi}_0\mathrm{d}\phi\,\frac{(\hat{n}'-\hat{n})_i\,(\hat{n}'-\hat{n})_j}{16\sin^4(\theta/2)} = \nonumber\\
    &\qquad\qquad\qquad\quad \frac{(\delta_{ij}-\hat{n}_i\hat{n}_j)}{2\pi}\log\left(\frac{\sin(\theta_{\rm max}/2)}{\sin(\theta_{\rm min}/2)}\right)\,. 
\end{align}
Here in the second integral, we have discarded terms $\sim \cos(\theta_{\rm max}) - \cos(\theta_{\rm min})$ on account of our small angle approximation. The other two integrals associated with the interference term between the $t$ and $u$ channels ($\propto \delta_{ab}|\hat{\bm n}'-\hat{\bm n}|^{-2}|\hat{\bm n}' + \hat{\bm n}|^{-2}$), end up giving contributions that go like $\sim \cos(\theta_{\rm max}) - \cos(\theta_{\rm min})$ and $\sim \log(\cos(\theta_{\rm min}/2)/\cos(\theta_{\rm max}/2))$. Under small angle approximation and large log, these become negligible and hence we discard them as well.

Informed by the target simulation system, we shall set the above Coulomb logarithm to be equal to $\log(m \sigma L) \equiv \Lambda$, where $L$ is the size of the system/simulation box size, $m$ is the lightest boson mass in the problem, and $\sigma$ is the velocity dispersion. With the above integrals, we get the following wave kinetic equation in the eikonal approximation 
\begin{align}
        &\frac{\partial f^a_{{\bm v}_a}}{\partial t} = \sum_{b}m_b^3\int\frac{\mathrm{d}\Omega_n}{4\pi}\frac{\mathrm{d}u}{2\pi^2}\Biggl[\frac{(4\pi m_am_bG)^2}{\mu}\Biggr]\frac{\Lambda}{4\pi}\;\times\nonumber\\
        &\qquad\Biggl[-\frac{1}{m_a}f^b_{\tilde{{\bm v}}_b}f^b_{\tilde{{\bm v}}_b}\,\hat{\bm n}\cdot\nabla_{{\bm v}_a}f^a_{{\bm v}_a} + \frac{1}{m_b}f^a_{{\bm v}_a}f^a_{{\bm v}_a}\,\hat{\bm n}\cdot\nabla_{\tilde{{\bm v}}_b}f^b_{\tilde{{\bm v}}_b}\nonumber\\
        &\qquad\; - \frac{\mu u}{m_am_b}(f^a_{{\bm v}_a} + f^b_{\tilde{{\bm v}}_b})(\delta_{ij}-\hat{n}_i\hat{n}_j)(\nabla_{v^i_a}f^a_{{\bm v}_a})(\nabla_{\tilde{v}^j_b}f^b_{\tilde{{\bm v}}_b})\nonumber\\
        &\qquad\; + \frac{\mu u}{2m_a^2}f^b_{\tilde{{\bm v}}_b}f^b_{\tilde{{\bm v}}_b}(\delta_{ij}-\hat{n}_i\hat{n}_j)\nabla_{v^i_a}\nabla_{v^j_a}f^a_{{\bm v}_a}\nonumber\\
        &\qquad\; + \frac{\mu u}{2m_b^2}f^a_{{\bm v}_a}f^a_{{\bm v}_a}(\delta_{ij}-\hat{n}_i\hat{n}_j)\nabla_{\tilde{v}^i_b}\nabla_{\tilde{v}^j_b}f^b_{\tilde{{\bm v}}_b}\Biggr]\,.
\end{align}
Now, redefining the relative velocity back to ${\bm u} = {\bm v}_a - \tilde{{\bm v}}_b$ so that the integration variable is $\tilde{{\bm v}}_b$, together with integration by parts (along with dropping boundary terms), and finally using the identities $\nabla_{x^i}[(|({\bm x}-{\bm y})|^2\delta_{ij} - (x^i-y^i)(x^j-y^j))/|({\bm x}-{\bm y})|^3] = -2(x^j-y^j)/|({\bm x}-{\bm y})|^3$, and $\nabla_{x}\cdot[({\bm x}-{\bm y})/|({\bm x}-{\bm y})|^3] = 4\pi\delta^{(3)}({\bm x}-{\bm y})$, we get Eq.~\eqref{eq:Boltzmann_gravity_eikonal} presented in the main text.\\

In summary, starting with the general \schr equation for $2$ body interactions, we first derived the full wave-kinetic Boltzmann equation~\eqref{eq:fdot_multicomponent_general} under the random phase approximation, which takes the form of Eq.~\eqref{eq:Boltzmann_gravity} for gravitational (or in general long range) interaction. For such long range interactions, the dominant contribution to the differential cross section comes from small angle wave scatterings. Suitably then, under the eikonal/small angle approximation, we derived the Fokker-Planck Eq.~\eqref{eq:Boltzmann_gravity_eikonal}. In the process, we have highlighted the presence of an interference term in the differential cross-section, readily interpreted as an interference between the $t$ and $u$ channels. Although its contribution is negligible under the eikonal approximation suited for long-range interactions, it may be important for other (e.g. short range) interactions. For example $T^{a,b,c,d}_{\bm k, \bm p, \bm q, \bm \ell} \propto \lambda_{ab}\,m_{a}m_{b}\,\delta_{bc}\delta_{da}$ in Eq.~\eqref{eq:fdot_multicomponent_general}, for point like interactions.

Also note that the Fokker-Planck equation~\eqref{eq:Boltzmann_gravity_eikonal} is identical to $f \gg 1$ limit of the quantum Boltzmann/Landau equation for bosons with long range interactions (under the small angle approximation). For instance see~\cite{Chavanis:2020upb}, for the relevant Landau equation. However, for general wave-mechanical system, this $f \gg 1$ route via the quantum version is not needed. In arriving at \eqref{eq:Boltzmann_gravity}, we did not assume $f \gg 1$. This equation therefore, applies generally for wave-systems that satisfy Schr{\"o}dinger-like equation, and entails the phenomenon of condensation. For instance even in our simulations, and well before the onset of condensate formation, $f$ is at-most order unity (near vanishing momenta). In this sense, Bose condensation is a wave-mechanical effect.

\section{Numerical Simulations}\label{app:numerical_details}

For our numerical studies, we have performed more than $\sim 100$ simulations in total of the multicomponent SP system Eq.~\eqref{eq:SP_general}, both for spin-$s$ (scalar, vector and tensor case) with equal boson mass and density for each component, as well as the two component scalar case with different masses and densities. 

To perform our simulations, we used two different codes, one being Python based {\sf{i-SPin}} integrator~\cite{Jain:2022agt}, and another Matlab based developed by Philip Mocz (modified to include multicomponent \schr field). The data presented for the spin-$s$ case in this paper was generated using the Matlab code. However, we have performed equivalent simulations using {\sf{i-SPin}} and confirmed the validity of our results. On the other hand the data for two component different mass case was mostly produced using {\sf{i-SPin}}. We have confirmed that spin and particle number are conserved to machine precision in these codes. 

For {\bf initial conditions}, we have worked with Gaussian initial profiles for $|\Psi^a_{\bm k}|$ for each species (with random phases for each ${\bm k}$ mode). For the spin-$s$ cases, there are $2s+1$ complex numbers, $\epsilon_a$, with $\sum_a\epsilon_a^{\ast}\epsilon_a=1$. Assuming equipartition, we choose them using a radially symmetric distribution function in $\mathbb{R}_{4s+2}$ and normalize such that the sum of their squares add up to unity. That is, they lie on the $S_{4s+1}$ hypersurface. On the other hand for two scalar case, we choose each of the respective phases for the two components (for every ${\bm k}$ mode) separately. 

To perform trustworthy simulations of the kinetic emergence of gravitating condensates, we chose the parameters $dx$ (discretization length scale), $dt$ (discretization time scale), $L$ (total box size), and $\bar{\rho}$ (average mass density) appropriately such that the kinetic regime condition $\tau_{\rm gr}\gg 1/m\sigma^2$ where $\tau_{\rm gr} = m^3\sigma^6/(\bar{\rho}^2(4\pi G)^2\Lambda)$ is the condensation time scale is satisfied, and the dynamics of the different waves in the simulation box are captured appropriately. Working with Gaussian initial ansatz~\eqref{eq:Gaussian_initialansatz}, dictating typical velocities to be $\sigma$, we measure length in units of $1/(m \sigma)$, time in units of $1/(m \sigma^2)$, and mass density in units of $(m^2 \sigma^4)/(4\pi G)\sqrt{\Lambda}$ where $\Lambda = \log(m \sigma L)$ is the Coulomb logarithm. For the spin-$s$ case when all the components have the same mass, we set 
\begin{align}
    dx &= \epsilon/(m \sigma) \qquad\qquad\qquad\qquad\;\;\;{\rm with}\quad \epsilon = 0.24\nonumber\\
    L &= \gamma/(m \sigma) \qquad\qquad\qquad\qquad\;\;{\rm with}\quad \gamma = 31\nonumber\\ 
    dt &= (2\pi/3) dx^2 m/\eta \qquad\qquad\quad\;\;{\rm with}\quad \eta = 1.5\nonumber\\
    \bar{\rho} &= \sqrt{\delta} m^2 \sigma^4/(4\pi G \Lambda^{1/2}) \  \qquad\;{\rm with}\quad \delta = 3.6\times 10^{-3}\,.
\end{align}
With the above parameters, we have performed $\sim 15$ simulations each for all of the scalar, vector, and tensor cases statistically similar initial conditions. Every simulation was run up until the threshold $\tilde{\rho} \sim 1$ was reached. These simulations were typically carried out at $N^3=128^3$, but we also checked individual cases with $N^3=256^3$ and found no discernable change in the condensation time. Contrary to changing the resolution, we also increased the box size to infer any IR effects. Upon doubling the box, we saw faster emergence of a halo like region, due to the Jeans instability scale associated with typical mass lumps in the box being smaller than the size of the box. Upon further evolution, we observed emergence of soliton within such halos. This was also seen in~\cite{Levkov:2018kau}.

As mentioned in the main body of the text, we define the condensation time as the time when the maximum density vs. time data points show a distinct change in slope on a log-log plot. We also tried different methods including the use of density thresholds, changes in running averages, linear regression of the slopes etc. These all yield qualitatively similar results.

For the unequal mass case, we have the same conditions as above, appropriately modified to accommodate shorter length scales and faster time scales associated with the heavier mass. Calling the smaller and heavier masses as $m_1$ and $m_2$ respectively with $m_2 = \{1, 1.5, 2\} \times m_1$, we set
\begin{align}
    dx &= \epsilon/(m_2 \sigma) \qquad\qquad\;\;\;{\rm with}\quad \epsilon = \{0.12, 0.19, 0.25\}\nonumber\\
    L &= \gamma/(m_1 \sigma) \qquad\qquad\qquad\quad {\rm with}\quad \gamma = 24\nonumber\\ 
    dt &= (2\pi/3) dx^2 m_1/\eta \qquad\qquad {\rm with}\quad \eta = 2\nonumber\\
    \bar{\rho} &= \sqrt{\delta} m_1^2 \sigma^4/(4\pi G \Lambda^{1/2}) \quad\quad{\rm with}\quad \delta \leq 5.1\times 10^{-3}\,.
\end{align}
Here $\bar{\rho} = \bar{\rho}_1$ is the average mass density of the first component. We carried out a total of $
\sim 50$ simulations with different mass and density ratios, different initial seeds, and $3$ different $\bar{\rho}_1$. Most of the simulations were carried out at $N^3=192^3$, with some smaller simulations at $128^3$. Again, no significant difference in condensation time was seen. \\ 

In some of our simulations (especially two component scalar case), we also kept track of the occupation number functions $f^a_{{\bm v}_a}$ of both components. For most of the simulations, we worked with Gaussian initial conditions (c.f. Eq.~\eqref{eq:Gaussian_initialansatz}) for which we find that for the component within which a condensate nucleates, its occupation number function develops increasing support towards smaller ``velocities", before eventually dropping at the onset of condensation nucleation. This conforms with our analytical understanding of the Boltzmann/Fokker-Planck equation~\eqref{eq:Boltzmann_gravity_eikonal}, as discussed in the main text. To test the validity of our understanding that the nucleation of condensate is characterized by small velocities, we also analyzed what happens with Dirac Delta initial distribution, i.e. $f^a_{{\bm v}_a} \propto \delta(|{\bm v}_a|-\sigma)$. Indeed, we find that $f^a$, for the species that forms the condensate, broadens out from the initial delta distribution and starts to develop increasing support over small velocities as time progresses. Eventually, the support drops, marking the nucleation of a condensate.

\end{document}